\documentclass[11pt]{article}

\usepackage[margin=1.25in]{geometry}
\usepackage[font=small,labelfont=bf,margin=0.4in]{caption}

\usepackage{amsmath,amssymb,amsthm}
\usepackage{bm}
\usepackage{graphicx}
\usepackage{float}
\usepackage{setspace}

\newcommand{\Z}{\mathbb{Z}}

\newcommand{\R}{\mathbb{R}}

\newcommand{\Var}{\mathrm{Var}}

\newcommand{\matern}{Mat\'ern}
\renewcommand{\S}{\mathbb{S}}
\newcommand{\wt}{\widetilde}
\newcommand{\sY}{\mathcal{Y}}
\newcommand{\sZ}{\mathcal{Z}}
\newcommand{\sU}{\mathcal{U}}

\usepackage[round]{natbib}
\usepackage{authblk}
\usepackage{color}

\title{Compression and Conditional Emulation \\ of Climate Model Output}

\begin{document}


\begin{center}

{\LARGE Compression and Conditional Emulation}
\vspace{3mm}

{\LARGE of Climate Model Output}
\vspace{3mm}

Joseph Guinness\footnotemark[1] and Dorit Hammerling\footnotemark[2]
\vspace{3mm}

\footnotemark[1]\small \textit{North Carolina State University, Department of Statistics}

\footnotemark[2]\small\textit{National Center for Atmospheric Research}
\end{center}

\footnotetext[1]{E-mail: jsguinne@ncsu.edu}
\footnotetext[2]{E-mail: dorith@ucar.edu}

\abstract{Numerical climate model simulations run at high spatial and temporal resolutions generate massive quantities of data. As our computing capabilities continue to increase, storing all of the data is not sustainable, and thus is it important to develop methods for representing the full datasets by smaller compressed versions. We propose a statistical compression and decompression algorithm based on storing a set of summary statistics as well as a statistical model describing the conditional distribution of the full dataset given the summary statistics. We decompress the data by computing conditional expectations and conditional simulations from the model given the summary statistics. Conditional expectations represent our best estimate of the original data but are subject to oversmoothing in space and time. Conditional simulations introduce realistic small-scale noise so that the decompressed fields are neither too smooth nor too rough compared with the original data. Considerable attention is paid to accurately modeling the original dataset--one year of daily mean temperature data--particularly with regard to the inherent spatial nonstationarity in global fields, and to determining the statistics to be stored, so that the variation in the original data can be closely captured, while allowing for fast decompression and conditional emulation on modest computers.}

\noindent \textbf{Keywords}: Spatial-Temporal Data, Gaussian Process, Half-Spectral, Nonstationary, SPDE

\section{Introduction}

The development of high performance computing facilities with ever-increasing power allows climate scientists to resolve and study small-scale climate and weather phenomena with high resolution model simulations. The resulting deluge of climate data is now a serious challenge for the advancement of climate science \citep{kunkel2014}. Climate model data are routinely discarded or not saved at desired resolutions to fit within storage constraints. The problem is especially pertinent for large multi-model ensemble projects such as the Coupled Model Intercomparison Projects (CMIP) used for the Intergovernmental Panel on Climate Change (IPCC) reports. CMIP5 storage requirements were approximately 2.5 PB, and an anticipated storage requirement of more than 10 PB is expected for the upcoming CMIP6 ensemble \citep{paul2015}. Similar issues arise with projects investigating internal variability, such as the Community Earth System Model (CESM) large ensemble project (CESM-LE) \citep{kay2015}, where the initial 30 ensemble members alone generated data exceeding 300 TB.

Compression methods are classically categorized as lossless, implying exact reconstruction of the data, or lossy, implying some loss of information \citep{sayood2012introduction}. The popular \textit{gzip} method, for example, is a lossless method. While the exact preservation of information is an attractive feature, lossless methods have proven ineffective for many scientific applications due to the random nature of the trailing digits of the floating-point data (\citet{lindstrom06}, \citet{bicer13}). Lossy compression algorithms, such as JPEG, wavelet compression, and compressed sensing \citep{donoho2006compressed,candes2006stable}, typically work by projecting the data onto a known basis and storing only the largest coefficients. Such methods are fast and have been shown to work well for image data, but they implicitly assume sparsity with respect to the basis and do not necessarily take advantage of the spherical space-time geometry or the persistent geographical features specific to climate data. Appendix \ref{basismethods} contains a comparison between our proposed methods and several basis function methods.

The application of lossy data compression to climate data is still in its infancy, so to obtain a better understanding of its effects, \citet{baker2014} investigated lossy compression in the context of ensemble variability and determined that a compression ratio of 5:1 produced acceptable results for yearly-averaged variables. The best results came from the \textit{fpzip} algorithm \citep{lindstrom06}, which truncates a specified number of least significant bits during the conversion of floating-point values to integers. One drawback of \textit{fpzip}, along with all other deterministic lossy compression algorithms, is the lack of uncertainty measures provided with the compressed data.

Recently, there has been interest in fitting statistical models to climate model output for the purpose of building computationally efficient model emulators \citep{castruccio2014statistical,castruccio2013global,williamson2014evolving,holden2015emulation,tran2016building} and for use as a tool for compressing the output. Previous work has viewed the parameters in the statistical model as the compressed object \citep{castruccio2015compressing,castruccio2015evolutionary}, and thus if the model is well-specified, it can be used to generate emulated model runs whose mean and covariance structure matches that of the original dataset. However, these methods are not designed to reconstruct individual model runs, which is of interest for compressing large ensemble projects that investigate internal variability in the climate models. Our work is focused simultaneously on accurately reconstructing the data and on accurately modeling its mean and covariance structure. In our approach, we store a set of summary statistics $S(Y)$ of the data $Y$, where the dimension of $S(Y)$ is smaller than that of $Y$. We estimate a statistical model $f(Y|S(Y))$ for the conditional distribution of the data given the summary statistics. Decompression is performed by either computing $E(Y|S(Y))$, or by taking a draw from $f(Y|S(Y))$. The compression ratio is defined as the ratio of the dimension of $Y$ to the sum of the dimension of $S(Y)$ and the number of parameters required to represent $f(Y|S(Y))$.

Since emulation refers to replacing the original data with a draw from a (unconditional) probability distribution $f(Y)$, we use the term ``conditional emulation'' to refer to taking a draw from $f(Y|S(Y))$. Having access to a statistical model that fits the data well is important because $E(Y|S(Y))$ is prone to oversmoothing in space and time, especially at high compression ratios. Using conditional draws allows us to closely match the amount of small-scale variation in the orginal data while exactly representing important features of the data summarized in $S(Y)$. This hybrid approach is highly flexible in that one can select which and how many statistics to store in order to target a close representation of particular features of the original data. Furthermore, storing summary statistics can act as a buffer against model misspecification; one can store portions of the data that are difficult to represent with a statistical model, for example non-Gaussian behavior or discontinuities along land/ocean boundaries. This setup implies that we need an accurate model for the conditional distribution of what is \emph{not} stored given what is stored; it is not necessary, however, to have a model for the marginal distribution of the stored statistics.

We consider the compression and decompression of one year of daily mean
temperatures, totaling 19.97 million values. In order to accurately model the major features of the data, we introduce a nonstationary space-time covariance model on the sphere that allows the characteristics of the time series to vary across the globe. The model is an instance of so-called {\it half spectral} models \citep{stein2005statistical}, in which the time series models are expressed in the spectral domain, and the coherence among the multiple time series is captured with a spatial covariance function whose parameters depend on frequency. This model specification allows us to overcome the computational burden of computing Gaussian likelihoods because maps of Fourier coefficients can be considered approximately uncorrelated across frequencies. Further, we employ a stochastic partial differential equation (SPDE) representation \citep{lindgren2011explicit} in the model for the Fourier coefficient maps, inducing sparsity in the inverse of the covariance matrix, which can be exploited to efficiently factor the matrix. This is the first time a half spectral model has been combined with a computationally efficient representation for the Fourier coefficients, which allows for the analysis of truly massive space-time datasets with Gaussian process models.

We consider Fourier coefficients as summary statistics. This choice is motivated both by the computational advantanges discussed above, and by the fact that for many locations, especially for the ocean pixels, the variation across time can be captured by a few low-frequency coefficients. Models are fit using conditional likelihoods. We introduce a method for automatically selecting the Fourier coefficients based on a greedy selection algorithm, and we propose a variant of this algorithm that can be distributed and is hence computationally efficient for high performance computing infrastructures typically available at climate data centers. We demonstrate that decompression can be performed quickly on a modest computer, having an end user with a laptop in mind. Finally, we evaluate the ability of the conditional simulations to accurately reflect the statistical properties of the small-scale features in the original data, and we conclude with a discussion.

\section{The Data}\label{datasection}

The CESM-LE project provides publicly available data from 40 ensemble model runs spanning the time period 1920-2100. We consider compression and decompression of one year (2081) of daily mean temperature fields from one ensemble member. The temperature values are reported on a $190 \times 288$ latitude $\times$ longitude grid, giving $n=54,720$ temperature values each day, and thus the entire year of daily data has 19.97 million observations.

\begin{figure}
\centering
\includegraphics[width=0.6\textwidth]{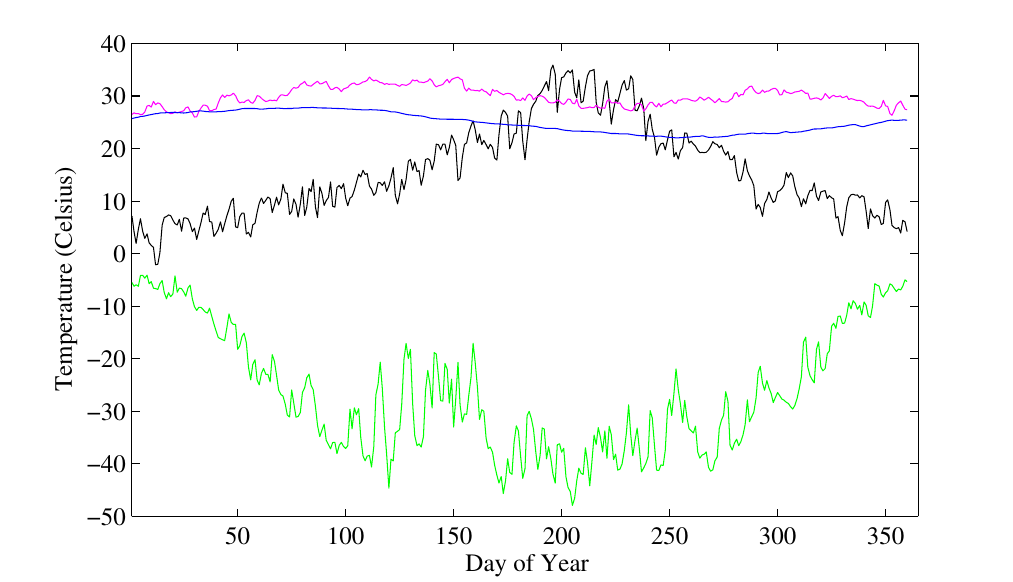}
\caption{Temperature time series from Chicago (black), Mumbai (magenta), south Atlantic Ocean (blue), and Ross Island, Antarctica (green).}
\label{timeseries}
\end{figure}

Let $Y(x,t)$ denote the daily mean temperature (in Celsius) at spatial location $x$ and day $t \in 1,\ldots,T = 365$. It is clear that the first- and second-order properties of the time series vary substantially around the globe. To see this, we plot in Figure \ref{timeseries} time series from pixels nearest Chicago, Mumbai, the southern Atlantic Ocean, and Ross Island in Antarctica. Visual inspection of the time series indicates that the means, the seasonal cycles, the variances, and the autocorrelations are quite different in the four series. Such characteristics of the time series can be studied by examining their Fourier coefficients. The discrete Fourier transform (DFT) over time at each pixel is given by
\begin{align*}
\sY(\omega_k; x) := \frac{1}{\sqrt{T}} \sum_{t=1}^{T} Y(x,t) \exp(-i\omega_k t),
\end{align*}
where $\omega_k = 2\pi k /T$ are the Fourier frequencies associated with $T$. These can be computed for all Fourier frequencies in $O(T\log T)$ floating point operations (flops) with fast Fourier transform (FFT) algorithms, and thus all of the DFTs at the $n = 190\times 288=54,720$ spatial locations can be computed in $O(n T\log T)$ flops. Throughout this work, we use the fft() function in Matlab to compute DFTs, which calls the FFTW library \citep{FFTW05}. To put it more concretely, all $n$ DFTs are computed in a total of 0.54 seconds on a Macbook Air with a 1.7GHz Intel Core-i7 processor and 8GB RAM on Matlab R2016b. We refer to this computer as our ``modest laptop.''

\begin{figure}
\centering
\includegraphics[width=\textwidth]{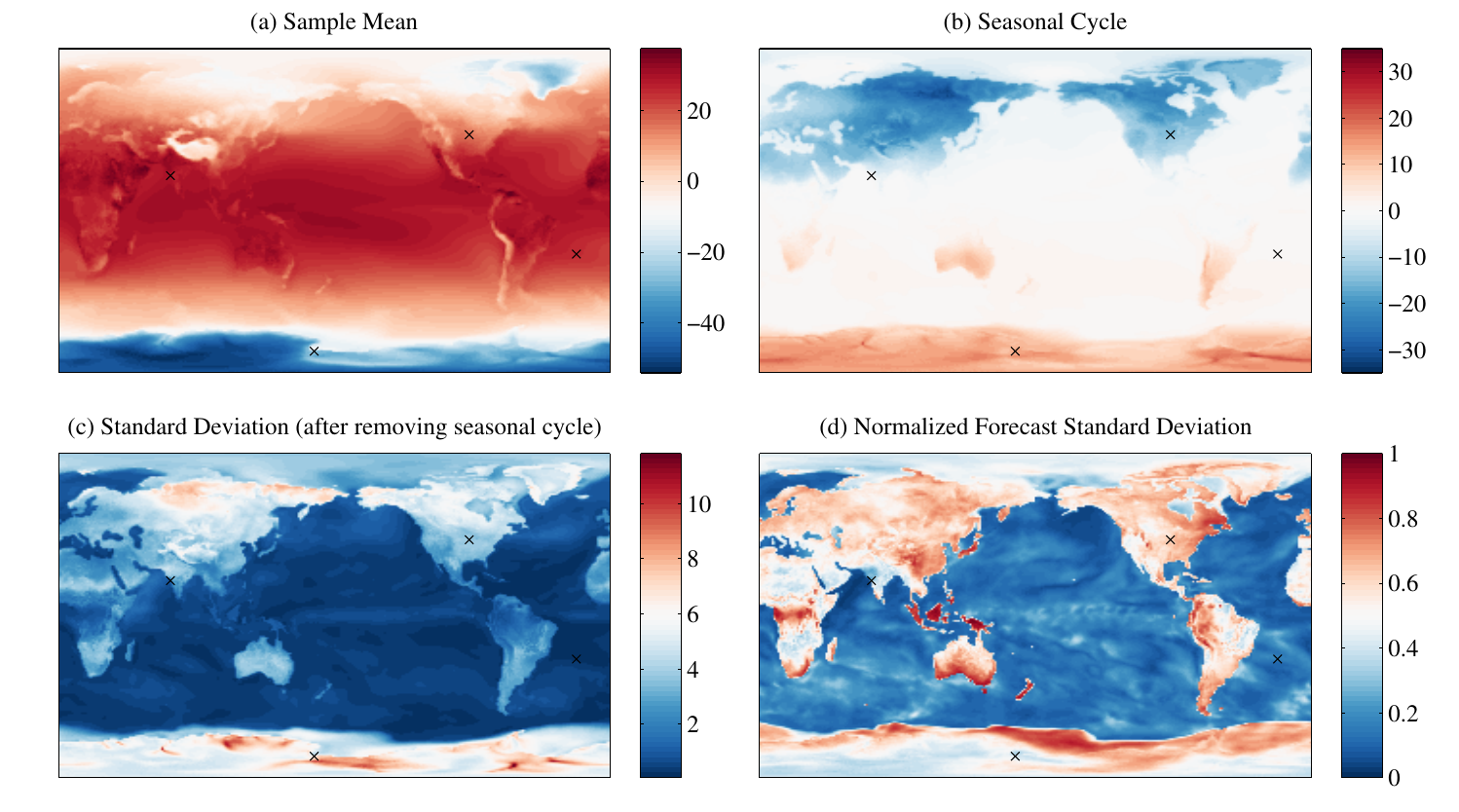}
\caption{Maps of sample mean, seasonal cycle, standard deviation of deseasonalized data (all in Celsius), and normalized forecast standard deviation (unitless). The black crosses indicate the locations of the time series plotted in Figure \ref{timeseries}.}
\label{summarymaps}
\end{figure}

Some of the major features of the spatial-temporal variation in the data can be summarized by making spatial maps of the Fourier coefficients $\sY(\omega_k;x)$, or functions thereof. In Figure \ref{summarymaps}(a), we plot a map of $T^{-1/2}\sY(\omega_0; x)$, which is simply the sample mean over time at each spatial location. In Figure \ref{summarymaps}(b), we plot the real part of $2T^{-1/2}\sY(\omega_1;x)$, which is variation in the data explained by $\cos(2\pi t/T)$, so it can be viewed as a signed strength of seasonal cycle--negative in the northern hemisphere and positive in the southern hemisphere. In Figure \ref{summarymaps}(c), we plot \begin{align*}
\widetilde{\sigma}(x) = \sqrt{\frac{1}{T-3} \sum_{k=2}^{T-2} |\sY(\omega_k; x)|^2},
\end{align*}
which is equal to the sample standard deviation of the deseasonalized time series.

If $Y(x,t)$ is modeled as a stationary process in time with spectral density $f(\omega; x)$, we can use the Fourier coefficients to form a nonparametric estimate of the spectral densities of the time series, as in
\begin{align*}
\widetilde{f}(\omega_k; x) = \frac{1}{2\pi}\sum_{j=0}^{T-1} \alpha(k-j)|\sY(\omega_j; x)|^2,
\end{align*}
where $\alpha$ is a smoothing kernel. In this exploratory data analysis, we use a Daniell window with bandwidth 17. Under the stationary-in-time assumption, Kolmogorov's formula \cite[Theorem 5.8.1]{brockwell2013time} gives the one-step-ahead prediction variance of $Y(x,t+1)$ based on an infinite past $Y(x,t),Y(x,t-1),\ldots$ in terms of the spectral density,
\begin{align*}
\Var(Y(x,t+1)|Y(x,t),Y(x,t-1),\ldots) = 2\pi \exp\left( \frac{1}{2\pi} \int_{0}^{2\pi} \log f(\omega; x) d\omega \right).
\end{align*}
In Figure \ref{summarymaps}(d), we plot a map of the estimated one-step-ahead prediction standard deviation normalized by the sample standard deviation,
\begin{align*}
\widetilde{\sigma}(x)^{-1}2\pi \exp\left( \frac{1}{T} \sum_{k=2}^{T-2} \log \widetilde{f}(\omega_k; x) \right)^{1/2}.
\end{align*}
Very smooth time series tend to have normalized one-step-ahead prediction variances near zero, and those with a value near 1 are likely to be well-modeled by white noise. Panels (c) and (d) indicate that the deseasonalized time series are generally more variable over land than over the oceans, and further, the time series are smoother over the oceans since their normalized forecast standard deviations are smaller as well, giving empirical evidence for the qualitative features seen in Figure \ref{timeseries}.

\begin{figure}
\centering
\includegraphics[width=\textwidth]{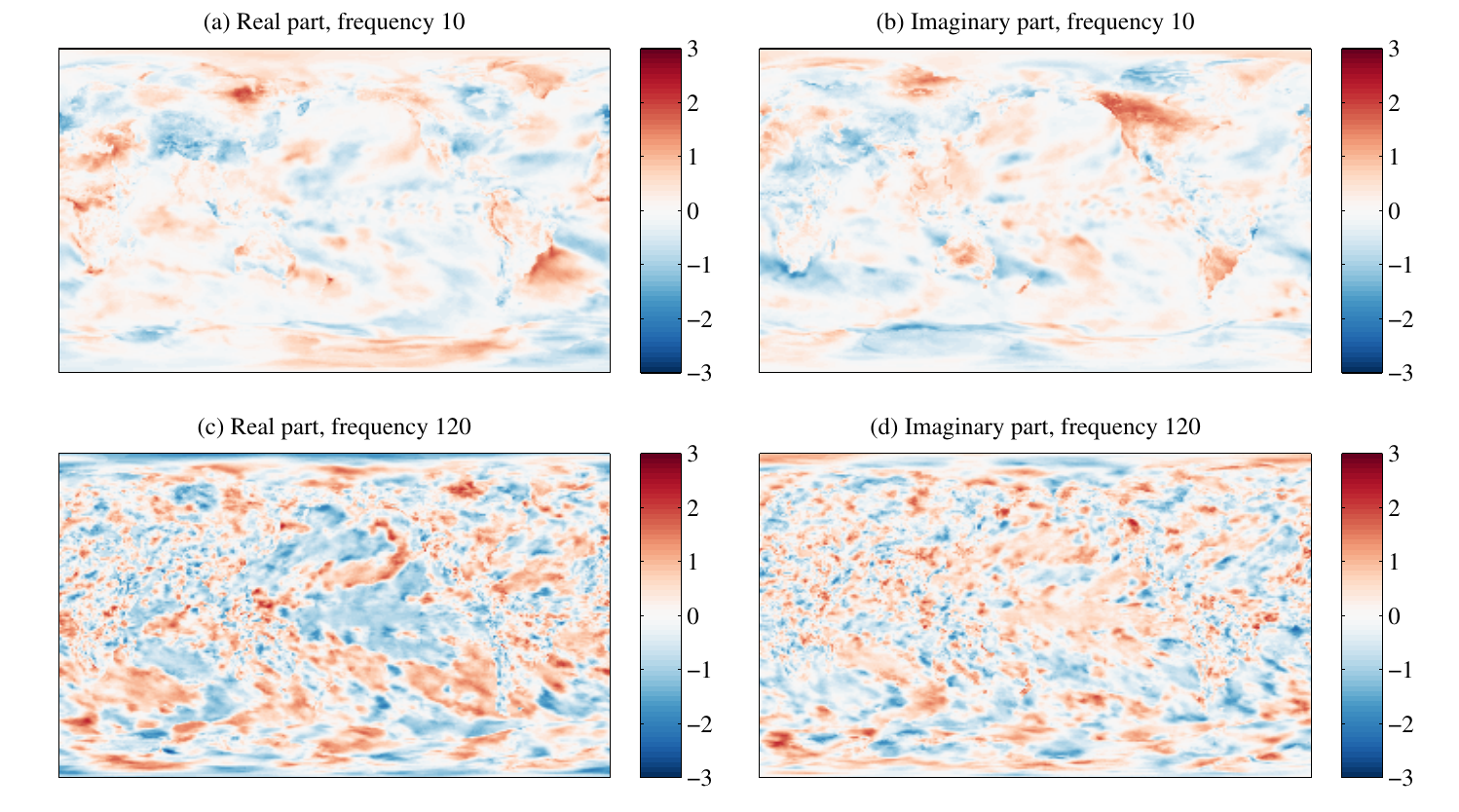}
\caption{Maps of the real and imaginary parts of $\sY(\omega_k; x)/\widetilde{f}(\omega_k; x)^{1/2}$, for $k = 10$ and $k = 120$.}
\label{Zplots}
\end{figure}

In Figure \ref{Zplots}(a-b) we plot the real and imaginary parts of $\sY(\omega_{10}; x)/\widetilde{f}(\omega_{10}; x)^{1/2}$, a map of low-frequency Fourier coefficients normalized by their estimated spectral densities, and in panels (c-d) we plot the real and imaginary parts of $\sY(\omega_{120}; x)/\widetilde{f}(\omega_{120}; x)^{1/2}$, a map of normalized high-frequency Fourier coefficients. We see that both the real and imaginary parts are each spatially correlated, and the spatial correlation is stronger at the lower frequency than it is at the higher frequency. This is consistent with the findings for wind data in \cite{stein2005statistical}, atmospheric pressure data in \cite{stein2009spatial}, and temperature data in \cite{guinness2013interpolation} and is also intuitive; low frequency fluctuations in temperature usually affect large regions simulataneously, whereas high frequency fluctuations can occur locally. The maps of normalized Fourier coefficients do not appear to deviate substantially from an assumption of isotropy. Possible exceptions include longer correlation ranges over the oceans and lack of dependence across land/ocean boundaries, especially at the lower frequency. In Section \ref{modelsection}, we present a way to model the spatial maps of scaled Fourier coefficients with spatial processes. The spatial model allows us to predict the full maps in Figure \ref{Zplots} from a subset of the pixels via computationally efficient spatial interpolation.

%
%

\section{Nonstationary Nonseparable Space-Time Model}\label{modelsection}

Let $\S^2$ be a sphere, and $\Z$ be the integers. We model temperature at location $x \in \S^2$ and time $t \in \Z$ as
\begin{align}\label{spacetimemodel}
Y(x,t) = \mu(x,t) + \frac{1}{\sqrt{2\pi}}\int_{0}^{2\pi} \exp(i\omega t) d\widetilde{\mathcal{Y}}(\omega,x),
\end{align}
where $\widetilde{\mathcal{Y}}(\omega,x)$ is an orthogonal increment process in $\omega$ for every $x$, with
\begin{align*}
E(d\widetilde{\mathcal{Y}}(\omega,x)d\widetilde{\mathcal{Y}}(\omega,y)^*) = \sqrt{f(\omega; x)f(\omega; y) }C(x,y; \omega)d\omega,
\end{align*}
where $*$ is complex conjugation, $C(x,y; \omega)$ is a coherence function, and thus $f(\omega; x)$ is the spectral density of the time series at location $x$. We assume $\mu(x,t)$ has the form
\begin{align*}
\mu(x,t) = \mu_0 + \mu_1\exp(i\omega_1 t) + \mu_1^* \exp(i\omega_{T-1} t),
\end{align*}
where $\mu_0$ is real, and $\mu_1$ is complex. The mean parameters capture the overall mean and seasonal cycle in the data.

Approximating the integral in \eqref{spacetimemodel} with a sum over $T$ Fourier frequencies,
\begin{align*}
Y(x,t) \approx \mu(x,t) + \frac{1}{\sqrt{T}}\sum_{k=0}^{T-1} \exp(i\omega_k t) \mathcal{Y}(\omega_k,x),
\end{align*}
along with the assumption that
\begin{align*}
E(\mathcal{Y}(\omega_k,x)\mathcal{Y}(\omega_j,y)^*) = \left\lbrace \begin{array}{ll}
\sqrt{f(\omega_k; x)f(\omega_k; y) }C(x,y; \omega_k) &\mbox{ if } k=j\\
0 & \mbox{ if } k\neq j,
\end{array} \right.
\end{align*}
leads to the usual approximation in the Whittle likelihood for multivariate time series \citep{whittle1953analysis} and affords large computational savings because the FFTs can be used for the DFTs, and the maps of Fourier coefficients $\sY(\omega_k;x)$ can be considered uncorrelated across frequencies when computing likelihoods and simulations. Computational details are given in Subsections \ref{conditionalloglikelihoods} and \ref{decompression}. Use of this approximation requires regularly spaced observations in time.

The model in \eqref{spacetimemodel} is a spatially nonstationary extension of the stationary model introduced in \cite{stein2005statistical}, which has been termed a ``half-spectral'' space-time model \citep{horrell2015half} due to the fact that the temporal covariances are expressed in the frequency domain. The principal advantage of modeling space-time data in this way is that the spatial maps of Fourier coefficients $\sY(\omega_k,x)$ are approximately uncorrelated across $\omega_k$. This brings enormous computational savings and allows us to flexibly specify and fit individual models to the spatial maps of Fourier coefficients at each frequency $\omega_k$, allowing for space-time nonseparability. Half spectral models have been applied to high-frequency space-time atmospheric pressure data \citep{stein2009spatial} and were generalized to handle high-frequency nonstationary-in-time temperature data \citep{guinness2013interpolation} with evolutionary spectra. Here, we allow the spectral densities $f(\omega,x)$ to vary with spatial location $x$, which gives nonstationary processes in space. There have been other proposals for modeling and computation with massive space-time datasets (e.g. \cite{katzfuss2012bayesian} and \cite{zhang2015full}), but none use the FFT to exploit regular spacing in time. \cite{guinness2015likelihood} use a three-dimensional FFT, which is applicable to data on a rectangular domain but not to data on a sphere.

The inclusion of nonstationarity in space is motivated by the time series plots in Figure \ref{timeseries}. In particular, our model allows the spectral densities to depend on spatial location, so that the temporal behavior is stationary in time at any given spatial location, but the temporal behavior is nonstationary in space. In order to specify a flexible set of time series models that are not overwhelmingly burdensome to store, we propose a semiparametric modeling approach. We express the spectral density for the time series at location $x$ as
\begin{align*}
f(\omega; x) = \exp \left( u_0(\omega) + \sum_{k=1}^K \theta_k(x)u_k(\omega) \right),
\end{align*}
where $\bm{\theta}(x) = (\theta_1(x),\ldots,\theta_K(x))$ are unknown parameters associated with each spatial location $x$, and $\bm{u}(\omega) = (u_0(\omega),u_1(\omega),\ldots,u_K(\omega))$ are data-dependent components of the log spectra. The first component $u_0(\omega)$ is the log average periodogram, averaged over pixels,
\begin{align}\label{u0}
u_0(\omega) = \log \left( \frac{1}{n} \sum_{i=1}^n |\sY(\omega; x_i)|^2 \right).
\end{align}
The $K$ components $u_1, \ldots, u_K$ are the first $K$ principal components of the log of smoothed and normalized periodograms. Details of the normalization and smoothing are given in Appendix \ref{periodogramsmoothing}. Estimation details for $\bm{\theta}(x)$ are given in Section \ref{estimationsection}. Our sensitivity analysis, presented in Appendix \ref{sensitivityanalysis}, shows that setting $K=1$ gave the smallest compression errors. The first principal component explained 99.3\% of the variation in the log smoothed normalized periodograms.

We model the coherences as
\begin{align*}
C(x,y; \omega) = K(  \| x-y \|; \kappa(\omega) ),
\end{align*}
where $K$ is the \matern\ covariance function with variance and smoothness equal to 1, $\| \cdot \|$ denotes Euclidean distance in $\R^3$, and $\kappa(\omega)$ is an inverse range parameter that depends on frequency $\omega$. Allowing the inverse range parameter to vary with frequency gives nonseparable space-time covariances and is motivated by Figure \ref{Zplots}.  There is a one-to-one mapping from $\R^3$ Euclidean distance to spherical distance. \cite{guinness2016isotropic} argue that there is little, if anything, to be lost by using the Mat\'ern with Euclidean distance in models for processes on the sphere. The variance parameter is set to 1 because the overall variance of the process is absorbed into the spectral densities. The choice of smoothness equal to 1 is motivated by balancing fit to the data and computational feasibility. The maps of Fourier coefficients are not particularly smooth in space, so a small smoothness parameter is justified. An integer smoothness allows us to employ the stochastic partial differential equation (SPDE) representation for \matern\ fields introduced in \cite{lindgren2011explicit}, which induces sparsity in the inverse of the covariance matrix, and thus has computational benefits both for maximum likelihood estimation of $\kappa(\omega)$, but, perhaps more importantly, for allowing users to perform fast conditional simulations on modest computers, one of the central goals of this work. Computational details are given in Subsection \ref{conditionalloglikelihoods}. Alternatively, one could employ a different method for computing likelihoods and predictions for large spatial datasets. Of those available, Vecchia's approximation \citep{vecchia1988estimation} may be particularly useful because it allows conditional likelihoods to be approximated. This is the first instance in which the half-spectral model has been paired with a flexible and computationally efficient specification of the coherence functions, a pairing that allows for the analysis of truly massive space-time datasets with Gaussian process models. The SPDE formulation also allows for the analysis of large spatial datasets with irregularly situated locations.

The burden of storing the model parameters is: $3$ numbers to represent $\mu_0$ and $\mu_1$, $n$ numbers to represent $\bm{\theta}(x)$, $2T/2$ numbers to represent $\bm{u}(\omega)$, and $T/2$ numbers to represent $\kappa(\omega)$, and thus the total model storage burden is $3 + n + 3T/2$, which is roughly $(0.0028)nT$, or $0.28\%$ of the size of the original dataset.

%
%

\section{Model Estimation and Summary Statistic Selection}\label{estimationsection}

We consider a subset of the Fourier coefficients $\sY(\omega_k; x)$ as the summary statistics $S(Y)$. The reason for this choice is two-fold. First, the approximate independence across frequency of the Fourier coefficients greatly simplifies the conditional simulations given the stored Fourier coefficients; we can conditionally simulate the remaining Fourier coefficients at each frequency independently of those at other frequencies, and then transform back to the time-domain with an inverse discrete Fourier transform. Second, at many of the spatial locations, especially the ocean locations, most of the variance of the time series is concentrated on a few low-frequency Fourier coefficients. As a consequence, many of the time series can be efficiently compressed by storing a small number of low-frequency Fourier coefficients.

The total storage capacity is dictated by the size of the original dataset and the desired compression ratio. For example, at a 10:1 compression ratio, we can store a total of 1.997 million numbers, which must be split between storing Fourier coefficients and model parameters. Accounting for the burden of storing model parameters leaves us with 1.942 million numbers left for representing Fourier coefficients. The total number of Fourier coefficients we can store is governed further by the ratio of zero frequency to non-zero frequency coefficients that are stored, since the non-zero frequency coefficients are complex, and thus require storage of two numbers. A further complicating matter is that, in addition to storing the Fourier coefficient $\sY(\omega_k; x)$, we must store $k$ and $x$. We assume that each frequency-location pair $(k,x)$ requires 8 bits on average, which we think is a conservative estimate and can be achieved with a standard delta encoding method \citep{sayood2012introduction}. Noting that the total number of stored Fourier coefficients is a somewhat complicated but easily computable function of the model, the compression ratio, and the distribution of coefficients among frequencies, we denote the number of stored coefficients as $N$ to simplify the discussion below.

Our task then is to select a set of $N$ Fourier coefficients $S(Y) = (\sY(\omega_1; x_1),\ldots,\sY(\omega_N,x_N))$ and model parameters $\bm{\theta}(x)$ and $\kappa(\omega)$ in order to best represent the original data and to accurately model the conditional distribution of the data given the Fourier coefficients. We consider the conditional likelihood to be a natural criterion for selecting the Fourier coefficients and the model parameters. The conditional likelihood is equal to the predictive density of the unstored coefficients, and thus larger conditional likelihoods correspond to sharper predictive distributions since all of the densities are assumed to be Gaussian. Further, the joint density for $S(Y)$ is irrelevant for this problem because there is no need to have an accurate statistical model for the data we store. The conditional likelihood given the stored coefficients is also a strictly proper scoring rule for prediction \citep{gneiting2007strictly}.

While the conditional likelihood is a natural selection criterion, maximizing it over all choices of $N$ Fourier coefficients and model parameters is an astronomically large combinatorial optimization problem (19.97 million choose 1.942 million). Thus, some severe restrictions on the search algorithm are necessary in order to make the selection computationally feasible. First, we estimate the time series model parameters once and for all by individually maximizing the Whittle likelihood for $\bm{\theta}(x)$ for each $x$. This part of the estimation takes on the order of 15 minutes on our modest laptop. We denote the estimates of the spectral densities as $\widehat{f}(\omega_k; x)$. The estimation of $\kappa(\omega)$ and a greedy algorithm for selecting Fourier coefficients are described in Subsection \ref{greedysearch}.

%
%

\subsection{Conditional Loglikelihoods}\label{conditionalloglikelihoods}

Before outlining the form of the conditional loglikelihoods, we first describe our notational convention. For variable $\mathcal{U}(\omega_k;x)$, which could be for example $\sY(\omega_k;x)$ or $\sZ(\omega_k;x) = \sY(\omega_k; x)/\widehat{f}(\omega_k;x)^{1/2}$, define the vector
\begin{align*}
\bm{\sU}(\omega_k) := \Big(\sU(\omega_k; x_1),\ldots,\sU(\omega_k; x_{n})\Big).
\end{align*}
We partition $\bm{\sU}(\omega_k)$ as $(\bm{\sU}_1(\omega_k),\bm{\sU}_2(\omega_k))$ so that $\bm{\sU}_1(\omega_k)$ is the vector of variables at frequency $\omega_k$ we store, and $\bm{\sU}_2(\omega_k)$ is the vector of unstored variables at frequency $\omega_k$. Further, define $\widehat{\bm{\sU}}_2(\omega_k) = E( {\bm{\sU}}_2(\omega_k) | \bm{\sU}_1(\omega_k))$.

Under the model \eqref{spacetimemodel}, each element of $\bm{\sY}(\omega_k)$ is approximately uncorrelated with each element of $\bm{\sY}(\omega_j)$ for $k\neq j$, and
\begin{align*}
\bm{\sZ}(\omega_k) \sim CN(0, \Sigma(\kappa(\omega_k))),
\end{align*}
where $CN$ represents a complex normal distribution. We partition $\Sigma(\kappa(\omega_k))$ as
\begin{align}\label{partitionM}
\Sigma(\kappa(\omega_k)) = \begin{bmatrix}
\Sigma_{11}(\kappa(\omega_k)) & \Sigma_{12}(\kappa(\omega_k))\\
\Sigma_{21}(\kappa(\omega_k)) & \Sigma_{22}(\kappa(\omega_k))
\end{bmatrix},
\end{align}
where $\Sigma_{ij}(\kappa(\omega_k)) = E( \bm{\sZ}_i(\omega_k) \bm{\sZ}_j(\omega_k)^\dagger )$ and $\dagger$ is complex conjugate transpose.

Writing $Q(\kappa(\omega_k)) = \Sigma(\kappa(\omega_k))^{-1}$ and partitioning $Q(\kappa(\omega_k))$ as in \eqref{partitionM}, the conditional expectation vector and covariance matrix of $\bm{\sZ}_2(\omega_k)$ given $\bm{\sZ}_1(\omega_k)$ are
\begin{align}
E(\bm{\sZ}_2(\omega_k)|\bm{\sZ}_1(\omega_k)) = \widehat{\bm{\sZ}}_2(\omega_k) &= -Q_{22}(\kappa(\omega_k))^{-1} Q_{21}(\kappa(\omega_k)) \bm{\sZ}_1(\omega_k), \label{condexp} \\
\Var( \bm{\sZ}_2(\omega_k) | \bm{\sZ}_1(\omega_k) ) &= Q_{22}(\kappa(\omega_k))^{-1}. \label{condvar}
\end{align}
The conditional loglikelihood for $\kappa(\omega_k)$ is
\begin{align*}
CL_k(\kappa(\omega_k)) = -\frac{1}{2}\sum \log \widehat{f}(\omega_k; x) + \frac{1}{2}\log \det Q_{22}(\kappa(\omega_k)) - \frac{1}{2} (\bm{\sZ}_2-\widehat{\bm{\sZ}}_2)^\dagger Q_{22}(\kappa(\omega_k)) (\bm{\sZ}_2-\widehat{\bm{\sZ}}_2),
\end{align*}
where the first sum is over all locations $x$ whose coefficients are not stored and corresponds to the marginal variances of the unstored coefficients. Due to the approximately zero correlation between $\bm{\sZ}(\omega_k)$ and $\bm{\sZ}(\omega_j)$ for $j\neq k$,
each $\kappa(\omega_k)$ can be estimated separately and in parallel by maximizing $CL_k(\kappa(\omega_k))$. The matrix $Q_{22}(\kappa(\omega_k))$ can be factored quickly due to the sparsity induced by the SPDE approximation. We factor the matrix with a Cholesky decomposition after permuting the rows and columns with a symmetric approximate minimum degree permutation \citep{amestoy1996approximate}, implemented in the symamd() function in Matlab, which encourages sparsity in the Cholesky factor. All Cholesky decompositions use chol() in Matlab, which for sparse inputs calls CHOLMOD \citep{davis2006direct}.

%
%

\subsection{Greedy Selection of Fourier Coefficients}\label{greedysearch}

Before selecting any Fourier coefficients, we estimate $\kappa(\omega_k)$ at each frequency using the marginal likelihood for all observations. Let $\widehat{\kappa}_0(\omega_k)$ denote the initial estimates. At this stage, we add selected coefficients before starting the search. Namely, for reasons discussed in Section \ref{sec:discussion}, we start the search with a low resolution grid of coefficients at frequencies $\omega_0$ and $\omega_1$.

Define $\mathcal{R}(\omega_k;x) = \sY(\omega_k;x)-\widehat{\sY}(\omega_k;x)$, the residual for predicting Fourier coefficient $\sY(\omega_k;x)$ from the stored Fourier coefficients, and let $\omega_*$ be the frequency for which $\mathcal{D}(\omega_k) = \max_{x} |\mathcal{R}(\omega_k;x)|^2$ is largest. We select the $M$ locations for which $|\mathcal{R}(\omega_*;x)|^2$ is largest, subject to the constraint that no two locations are within distance $d_{\min}$ of each other. We add the Fourier coefficients for the selected locations to $\bm{\sY}_1(\omega_*)$ and recompute $\mathcal{R}(\omega_{*};x)$. We repeat this process until $N$ Fourier coefficients have been added. This iterative procedure is stopped periodically to re-estimate $\kappa(\omega_k)$ using the conditional loglikelihoods given the current selection of stored Fourier coefficients. We denote the estimated coherence parameters at the $j$th stop as $\widehat{\kappa}_j(\omega_k)$ and the total number of stops with $J$.

If $M = 1$, this method corresponds to sequentially picking the coefficient with the largest residual under the current conditional model, and thus the choice of $d_{\min}$ would not be relevant. Since each iteration requires factoring $Q_{22}(\kappa(\omega_k))$, the algorithm is faster when $M$ is larger because we add a large number of points at each iteration, so fewer total iterations are required. The choice of $d_{\min}$ should in principle depend on $M$ because if only a few locations are added per iteration (i.e. $M$ is small), we should ensure that no two points are close to each other by making $d_{\min}$ large.

This particular method is intentionally iterative so that the estimates of $\kappa(\omega_k)$ are allowed to change when additional coefficients are stored, which is important since the maps of Fourier coefficients exhibit some deviations from the assumption of isotropy. For example, since the coefficients are generally smoother over the oceans than over land, if we were to store all of the land coefficients at frequency $\omega_k$, the conditional loglikelihood estimates $\widehat{\kappa}_J(\omega_k)$ would be smaller than the marginal loglikelihood estimates $\widehat{\kappa}_0(\omega_k)$. Additionally, storing coefficients for which $|\mathcal{R}(\omega_k;x)|^2$ is large encourages selection of coefficients that are either far in distance from already stored coefficients, or whose conditional mean is not accurately represented in the conditional distribution, a sign of possible anisotropy near $x$.

We have also developed an alternative version of the algorithm, where $M$ locations are distributed to the frequencies $\omega_k$ proportional to $\mathcal{D}(\omega_k)$. The number of locations assigned to frequency $\omega_k$ is defined as
\begin{align*}
m_k := \frac{ \mathcal{D}(\omega_k)}{\sum_{k=0}^{T/2} \mathcal{D}(\omega_k)} M.
\end{align*}
The remaining steps are identical to the sequential version, where, for each frequency $\omega_k$,  the $m_k$ locations for which $|\mathcal{R}(\omega_k;x)|^2$ is largest are chosen, subject to the minimum distance constraint. Then $\mathcal{R}(\omega_{k};x)$ and $\mathcal{D}(\omega_{k})$ are recomputed. The periodic re-estimation of $\kappa(\omega_k)$  is also analogous. The main advantage of this ``distributed" variant of the search algorithm is the ability to parallelize it, which can provide considerable savings in computational time.  This is especially pertinent for lower compression ratios, for example 5:1, where a large number of locations are to be added.

We conducted a thorough analysis of the effect of various choices of $M$, $J$ and $d_{\min}$, both on the timing of the algorithms and on the compression results. The analysis in Section \ref{sec:comparison} presents results for $M=50$ (sequential) and $M=7049$ (distributed), which corresponds to  $m_k  \approx 38$ per frequency on average, although the actual selection is heavily weighted towards lower frequencies. For both algorithms we used $J=8$ and $d_{\min} = 0.05$. These settings provided decompressed datasets with small prediction errors and ran in a reasonable amount of time. This choice of $d_{\min}$ corresponds to a width of 2.3 pixels at the equator and guards against selecting too many pixels near the poles. Appendix C contains an analysis of the sensitivity of RMSPE with respect to $d_{\min}$.

\subsection{Decompression}\label{decompression}
Decompression is achieved by computing conditional expectations and conditional simulations given the stored coefficients. The conditional expectation of the scaled Fourier coefficients $\bm{\sZ}_2(\omega_k)$ given $\bm{\sZ}_1(\omega_k)$ can be computed as in Equation \eqref{condexp}. Then we compute $\widehat{\sY}(\omega_k; x) = \sqrt{f(\omega_k;x)} \widehat{\sZ}(\omega_k,x)$ for each $k$ and $x$. Finally, temperature values are obtained by computing inverse DFTs of $\widehat{\sY}(\omega_k;x)$ at each location. The most demanding computational task is computing the conditional expectations, which involves solving a linear system with $Q_{22}(\kappa(\omega_k))$, accomplished first by computing the Cholesky factorization $LL^T = Q_{22}(\kappa(\omega_k))$, which is feasible due to the sparsity of $Q_{22}(\kappa(\omega_k))$, and then by solving the lower triangular sytem $L v = -Q_{21}(\kappa(\omega_k))\bm{\sZ}_{1}(\omega_k)$, followed by solving the upper triangular system $L^T u = v$, and setting $E(\bm{\sZ}_2(\omega_k)|\bm{\sZ}_1(\omega_k)) = u$. We use the Cholesky factor method because the Cholesky factor can also be used for simulating the conditional residuals $e_2$. For this task, we solve the upper triangular system $L^{T} e_{2} = \varepsilon_2$, with $\varepsilon_2 \sim N(0,I)$, so that $e_2$ has covariance matrix $L^{-T} L^{-1} = Q_{22}^{-1}(\kappa(\omega_k))$.

The model has been chosen so that decompression can be performed quickly on a modest computer, while still allowing for substantial nonstationarity. Decompression can be parallelized as well by treating each frequency separately. The time required for decompression varies by strategy and the number of stored coefficients--storing more coefficients at frequency $\omega_k$ means that $Q_{22}(\kappa(\omega_k))$ is smaller and thus can be factored faster--but decompression generally took between 3 and 5 minutes without parallelization and between 1 and 2 minutes parallelizing over the 2 cores on our modest laptop.

\section{Comparisons Between Original and Decompressed Datasets} \label{sec:comparison}

\begin{figure}
\centering
\includegraphics[width=0.9\textwidth]{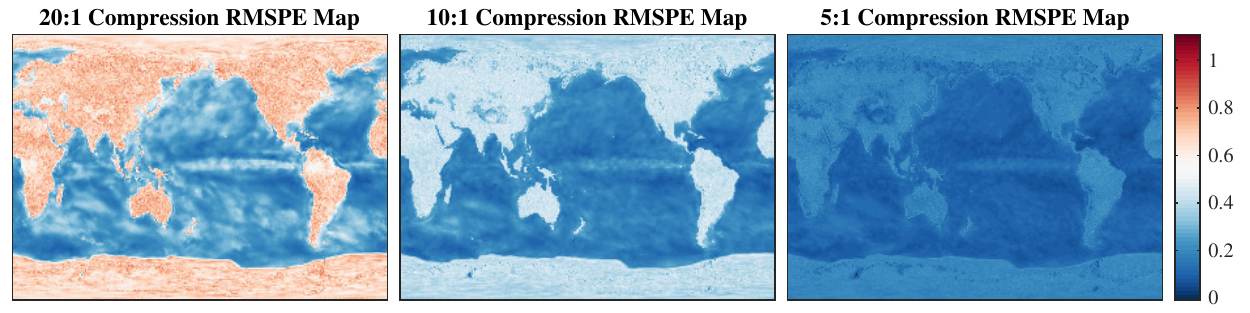}    
\caption{Maps of pixelwise RMSPE. Units are degrees Celsius.}
\label{rmsemaps}
\end{figure}

Let $\widehat{Y}(x,t)$ denote the conditional expectation given the stored Fourier coefficients, and $\wt{Y}(x,t)$ denote a realization of the conditionally simulated data. The first performance criterion we consider is a pixelwise root mean squared prediction error (RMSPE), defined as
\begin{align*}
s(x) = \sqrt{\frac{1}{T} \sum_{t=1}^T \Big(Y(x,t) - \widehat{Y}(x,t) \Big)^2 }.
\end{align*}
We plot maps of $s(x)$ for each level of compression in Figure \ref{rmsemaps}. In Table \ref{tab:results}, we report the square root of pixel-area-weighted averages of $s^2(x)$, averaged over water locations, land locations, and over all locations. As expected, the errors decrease as the compression ratios decrease, and the average errors over ocean pixels are smaller than over land pixels. There are minor differences between the two algorithms, notably that the sequential version performs generally better over the ocean. The RMSPE differences are more prominent for the higher compression ratios, where the gains in computing times are also proportionally lower, so for practical purposes the choice of the selection algorithm could be a function of the compression needs and the available computing resources. For both algorithms, the computing times increase sublinearly with the number of coefficients selected. This finding is as expected, as we left the number of times $\kappa(\omega)$ is re-estimated, which is computationally demanding, constant for all compression ratios. Reported computing times in Table \ref{tab:results} are from a single Geyser node on the Yellowstone Supercomputer at The National Center for Atmospheric Research. Geyser nodes have four 10-core 2.4GHz processors and 1TB of memory. The RMSPE values should be interpreted with reference to Figure \ref{summarymaps}, recalling that some of the deseasonalized time series in polar regions can be modeled as nearly white noise with standard deviations as large as 10 degrees Celsius, so a RMSPE of less than 1 degree Celsius is small in that context.
\begin{table} [H]
\centering
\begin{tabular}{c|c|ccc|c} \hline
Selection algorithm  & Comp. Ratio & land & ocean &  all & runtime\\ \hline
& 20:1 & 0.6864  &  0.2780  &  0.4367 & 4.33 \\
sequential & 10:1 &  0.4203  &  0.1888  &  0.2762  & 8.30 \\
& 5:1 & 0.1991   & 0.1150 &   0.1444      &14.13  \\  \hline
& 20:1 & 0.6937 &   0.3239 &  0.4618 &  1.15 \\
distributed & 10:1 & 0.4202  &  0.2153 &  0.2896 & 1.63 \\
 & 5:1 & 0.1971  &  0.1225  &  0.1479 & 2.32 \\ \hline
\end{tabular}
\caption{\small RMSPEs and run times for three different compression ratios and the two different greedy selection algorithms. Means are pixel area weighted averages. Runtimes are shown in hours on a single Geyser Yellowstone node with four 10-core 2.4GHz processors and 1TB memory. }
\label{tab:results}
\end{table}

\begin{figure}
\centering
\includegraphics[width=0.9\textwidth,angle=0]{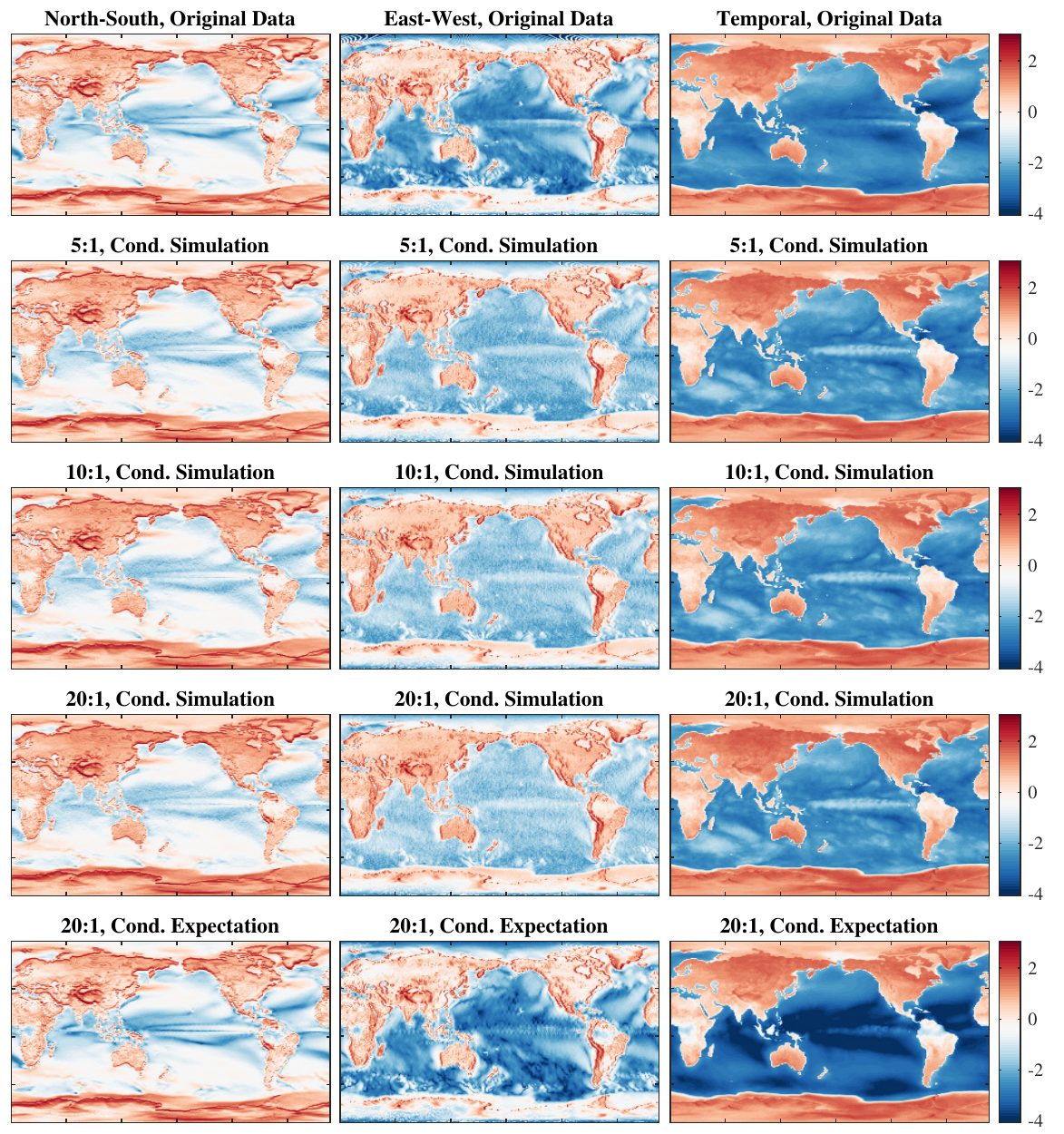}
\caption{Maps of log contrast variances. First column are average North-South contrast variances, middle column are east-west contrast variances, and third column are one-step temporal contrast variances. First row is computed from the original data, second through fourth rows from conditionally simulated data at the three compression levels, and the last row is from conditional expectation data at 20:1 compression ratio.}
\label{contrastall}
\end{figure}

In addition to producing small RMSPEs, the decompressed data also accurately capture the spatial-temporal dependence present in the original data. To see this, we plot spatial maps of averaged contrast variances, computed from the original data and from the conditionally simulated data. Define $\delta_{\mbox{lat}}$ and $\delta_{\mbox{lon}}$ to be displacement vectors of one pixel in latitude and longitude, respectively, so that $Y(x,t) - Y(x+\delta_{\mbox{lat}},t)$ is a one pixel north-south contrast, and likewise $Y(x,t) - Y(x+\delta_{\mbox{lon}},t)$ is a one pixel east-west contrast. In Figure \ref{contrastall}, we plot maps of the log of
\begin{align}\label{contrastequations}
\mbox{North-South}: \quad &{\frac{1}{T} \sum_{T=1}^T \Big(Y(x,t)-Y(x+\delta_{\mbox{lat}},t)\Big)^2},\\
\mbox{East-West}: \quad &{\frac{1}{T} \sum_{T=1}^T \Big(Y(x,t)-Y(x+\delta_{\mbox{lon}},t)\Big)^2},\\
\mbox{Temporal}: \quad &{\frac{1}{T-1} \sum_{T=1}^{T-1} \Big(Y(x,t)-Y(x,t+1)\Big)^2},
\end{align}
which are, respectively, the average north-south, east-west, and one step temporal contrast variances. Likewise, for each compression level, we plot the corresponding maps with $Y$ replaced by $\widetilde{Y}$ (conditional simulations). Although we did not introduce any anisotropy in the spatial dependence of the scaled Fourier coefficients, the contrast variances of the conditionally simulated data closely match those of the original data, with the accuracy increasing as the compression level decreases. Lastly, Figure \ref{contrastall} also includes a map of contrast variances with $Y$ replaced by $\widehat{Y}$ (conditional expectation) for the 20:1 compression level. It is evident that conditional expectation maps are much too smooth in time over the oceans and also slightly too smooth in space over some land pixels. This oversmoothing behavior is why it is important, especially at the highest compression ratios, to use conditional simulations in the decompression step. The plotted maps are from the sequential selection algorithm; the distributed algorithm produced visually indistinguishable results.

\begin{figure}
\centering
\includegraphics[width=0.9\textwidth]{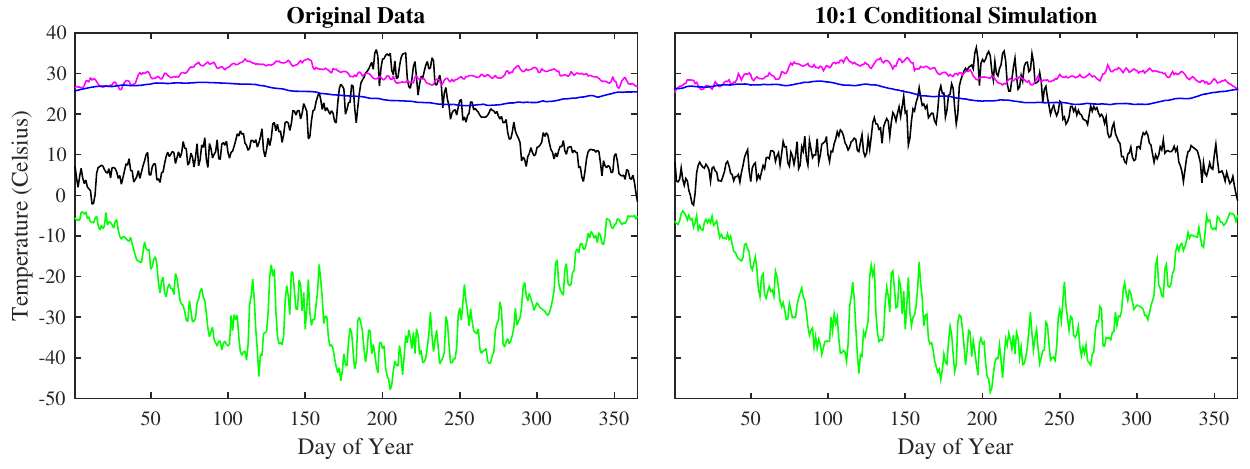}
\caption{Original and 10:1 decompressed time series plots from Chicago (black), Mumbai (magenta), south Atlantic Ocean (blue), and Ross Island, Antarctica (green).}
\label{timeseries_decompressed}
\end{figure}

In Figure \ref{timeseries_decompressed}, we reproduce the original time series plots from Figure \ref{timeseries}, as well as the time series from the conditionally simulated data at the 10:1 compression level. The conditional simulations are able to reproduce the temporal dependence structures that vary substantially over space in the original data. The decompressed data also recover much of the small-scale structure as well. We point out here that even though the time series models were assumed to be stationary, the decompressed data accurately reflect some of the temporal nonstationarity as well; for example, in the Ross Island time series, the variance is larger in the middle of the year, and the decompressed data preserve this feature even though the assumed model is stationary across time.

\section{Discussion}\label{sec:discussion}

{\it Model Assumptions}: The model allows the spectral densities governing the temporal dependence to vary over the globe, an obvious feature in the data, as seen in Figure \ref{timeseries}. The model assumes that the standardized coefficients $\sY(\omega_k,x)/f(\omega_k,x)^{1/2}$ form an isotropic Gaussian process over the globe.  Based on a visual inspection of Figure \ref{Zplots}, a case could be made that this assumption is violated in that the correlation ranges appear to be longer over the oceans, and exhibit other nonstationary or non-Gaussian features, especially at low frequencies. Future work could address these modeling issues, particularly nonstationary correlation in the Fourier coefficients. However, there is a subtle point to be made: isotropy is only assumed in determining the conditional distributions for the unsaved coefficients given the saved ones. This is why it is important to use the conditinal loglikelihoods to estimate the spatial coherence parameters. No distributional assumptions are made about the saved coefficients--they are what they are, so we can use our choice of the saved coefficients to preserve nonstationarity or non-Gaussianity when it exists. Our algorithm for selecting the Fourier coefficients appears to be preserving nonstationarity automatically, as we can see in Figure \ref{contrastall}--the East-West and North-South contrast variances of the conditionally simulated data closely match the nonstationarity in the original data. We can see this effect in the time series plots as well. There is some evidence in Figures \ref{timeseries} and \ref{timeseries_decompressed} that the time series are not stationary. For example, the Ross Island time series appears to have a larger variance in the winter months. Even though we modeled the time series as stationary processes, the conditionally simulated data in Figure \ref{timeseries_decompressed} appear to capture this nonstationary feature.

\vskip11pt

\noindent \textit{Low Frequency Coefficients}: At the lowest few frequencies, use of the conditional loglikelihood to estimate the spatial coherence parameters results in estimates of $\kappa(\omega)$ that correspond to processes that are visually rougher than the data. Since there is so much variation across the globe in the means and seasonal cycles, we hypothesize that departures from the Gaussian process assumption at those frequencies could be causing this behavior. A detailed investigation of non-Gaussian processes for the lowest frequencies is beyond the scope of this paper; however, we found that imposing some specific settings at the lowest frequencies improved the results. For $k=0,1$, we start the greedy search for coefficient selection by manually adding low resolution grids of 6,840 locations, which avoided selecting too many low frequency coefficients in early iterations of the distributed algorithm. For $k=0,1,2$, we set $\widehat{\kappa}(\omega_k) = 0.01$, which improved the contrast variance maps.

\vskip11pt

\noindent {\it Storing Numbers at a Fixed Precision}: The original data are stored as single precision floating point numbers. In calculating the compression ratios, we have assumed throughout that all parameters and Fourier coefficients are stored at the same precision, and that the frequency-location pairs are stored as an increasing sequence of integers to allow for delta encoding. Traditional compression algorithms make heavy use of storing different parts of the datasets at different precisions. We think that these ideas could certainly be applied to the present problem in conjunction with our proposed geostatistical compression methods, but we leave that problem for future work.

\vskip11pt

\noindent {\it Polar Regions}: The original data are reported on a regular latitude/longitude grid, which means that the original data devote a disproportionate amount of storage to the polar regions. An attractive feature of our methods is that the polar regions are treated in a natural fashion by the SPDE model, and our method for choosing the Fourier coefficients avoids selecting too many near the poles. Further, the SPDE mesh can be easily adapted to non-regular grids in latitude and longitude.

\vskip11pt

\noindent{\it Parameter Estimation and Inference}: We did not attempt to perform a global optimization over the spatial range parameters and the spectral density parameters. Instead, we first fit the spectral density parameters individually at each pixel and then fixed them when maximizing the conditional loglikelihoods over the spatial range parameters. Performing the global optimization is not tractable because there are $n$ time series parameters, and moreover, such a maximization could not be done separately at each frequency because the spectral density parameters affect the spectral densities of all frequencies. A different modeling strategy would be required for such global optimizations, possibly at the expense of having to assume a simpler, perhaps space-time separable, model. We also do not focus on model inference in this application. Compression differs from most prediction problems because we observe the data that we ultimately plan to predict at the decompression step. This distinction means that issues of model inference and overfitting do not hold the central importance that they do in more standard prediction problems.

\vskip11pt

\noindent{\it Fixing the Compression Ratio}: The compression ratio was fixed at three values, 20:1, 10:1, and 5:1, and then the model and coefficients were selected subject to the compression ratio. This framework is chosen for simplicity of presentation, and the methods we propose could be easily adapted, for example, to a setting where one desires to reach a specific outcome, such as a certain RMSPE or maximum absolute prediction error, with the best possible compression ratio.
\vskip11pt

\noindent{\it Other Climate Variables}: We chose temperature for this study because of its importance to climate science and because it presented us with a difficult modeling task due to its nonstationarity. Other variables pose modeling challenges that may be more or less difficult. For example, the Gaussian assumption is harder to defend for precipitation, ice cover, and sulphur burden, and so one may need to resort to non-Gaussian models for those variables. The development of non-Gaussian spatial-temporal models is an important and ongoing area of research in our field. We do think that our proposed general concept of storing a set of summary statistics and a conditional model is a useful strategy for compression, whether or not that model is Gaussian. We have also not investigated 3-D variables, which will require more research and new modeling approaches. Another avenue for research is statistical compression of multiple climate variables simultaneously, which would allow us to condition on multivariate statistics, which is important for respecting physical constraints. This approach requires modeling multivariate relationships among variables on a sphere, and there has been some work on this topic (e.g.\ \cite{jun2011non}).
\vskip11pt

\noindent{\it Computation and Scalability}: The code is written in Matlab version R 2016b. We include the Matlab code for reproducing the results in the online supplementary material. All the computations were executed on the Geyser nodes on the Yellowstone high performance computing system of the National Center for Atmospheric Research (NCAR). The timing results for a single year of data shown in Table \ref{tab:results} were obtained using one 40-core node. The new Cheyenne supercomputer managed by NCAR has over 145,000 cores, and so scaling to multiple years of data and multiple climate variables is computationally very feasible. While we have conducted some optimization and implemented the parallelization, if this method were to be applied in an operational mode, the code will be handed over to a software engineering team to further optimize some aspects of the code such as I/O and will likely be refactored in a compiled language, which is standard practice at large climate research centers.

\section*{Acknowledgements}\label{sec:ack}

We acknowledge high-performance computing support from Yellowstone (ark:/85065/d7wd3xhc) provided by NCAR's Computational and Information Systems Laboratory, sponsored by the National Science Foundation. We further acknowledge  support for long-term collaborative visits from the NSF Research Network on Statistics in the Atmosphere and Ocean Sciences (STATMOS) through grant DMS-1106862. This material is based upon work supported by the National Science Foundation under Grant Nos.\ 1406016 and 1613219. We would like to thank Allison Baker for sharing the data and helpful discussions and generous advice regarding its usage. We would like to thank Anthony Tracy, Sophia Chen and William Kaufman for executing some of the HPC runs involved in this research.

\appendix

\section{Principal Components of Smoothed Periodograms}\label{periodogramsmoothing}

The vector $\bm{u}_j = (u_j(\omega_0),\ldots,u_j(\omega_{T-1}))$ is defined as the $j$th principal component of
\begin{align*}
g(\omega_k,x) = \log\left( \sum_{\ell=0}^{T-1} \alpha(\ell - k) |\mathcal{Y}(\omega_\ell,x)|^2/\exp(u_0(\omega_\ell)) \right),
\end{align*}
where $u_0(\omega_\ell)$ is defined in \eqref{u0}, and $\alpha(\ell-k)$ is a smoothing kernel defined as
\begin{align*}
\alpha(\ell) = \frac{1}{c} \exp( 100 (\cos(\omega_\ell) - 1) ),
\end{align*}
where $c$ is a normalizing constant to ensure that $\alpha$ sums to 1.

\section{Basis Function Compression}\label{basismethods}

Traditional compression algorithms use a represention of the data as a linear combination of known basis functions and work by storing the largest coefficients in the basis function representation. For data on a sphere, a natural choice is a spherical harmonic basis, since the spherical harmonic basis function representation is the unique doubly orthogonal (i.e.\ orthogonal basis functions and uncorrelated coefficients) for isotropic processes on the sphere \citep[Theorem 1, page 73]{yadrenko1983spectral}. We also consider a two-dimensional wavelet basis and a basis consisting of the eigenvectors of the empirical spatial covariance matrix of the temperature data, also known as a principal components analysis (PCA).

The spherical harmonic and wavelet bases are mathematical functions and thus do not incur any storage cost. We use the spharm Matlab function \citep{spharm} to compute the spherical harmonic functions and the wavedec2 Matlab function for the wavelet projections. We obtained the best results with the sym4 Symlet wavelets, with level $N=4$. The PCs do incur a storage burden, since each basis function is a complete spatial map. Thus we can use a maximum of 18, 36, and 73 PCs for 20:1, 10:1, and 5:1 compression, respectively. The cost of storing the coefficients is negligible compared to the cost of storing the PCs. For all of the three methods, we project each day of temperature data onto the chosen basis and store the number of coefficients dictated by the compression ratio.

The resulting overall area-weighted RMSPEs are provided in Table \ref{tab:basisresults}. We also reproduce the sequential and distributed algorithm results from Table \ref{tab:results}. The spherical harmonics of large order were numerically linearly dependent, and thus we could not obtain results for the 5:1 compression level. We can see that none of the basis function methods are competitive with the methods described in this paper on the metric of RMSPE.

We also reiterate here that in addition to achieving small prediction errors, our methods are capable of fast conditional emulation in the context of an estimated statistical model, which as can be seen in Figure \ref{contrastall}, is important for reproducing the spatially-varying statistical characteristics of small scale noise. Extending this feature to basis function compression methods would require additional research.

\begin{table}
\centering
\begin{tabular}{cccc}
 & \multicolumn{3}{c}{ratio} \\
 \cline{2-4}
 Method & 20:1 & 10:1 & 5:1 \\
 \hline
Sequential & 0.4367 &0.2762  &0.1444  \\
 Distributed &0.4618 &0.2896 &0.1479 \\
 Spherical Harmonics & 1.3649 & 1.0568 & N/A \\
 Wavelets &0.9011 &0.5429 & 0.2620\\
 PCA & 1.5337 & 1.2089 & 0.8877
\end{tabular}
\caption{Area-weighted overall RMSPE for the sequential and distributed algorithms described in this paper, spherical harmonics method, wavelet method, and the EOF method.}
\label{tab:basisresults}
\end{table}

\section{Sensitivity Analysis}\label{sensitivityanalysis}

This appendix contains an analysis of the sensitivity of the results to some of the choices made in the model and the compression algorithm. We varied the choice for $\kappa(\omega_k)$ when $k=0,1$, or 2 at $0.01$, $0.1$, and $1.0$. The RMSPE values did not vary substantially among the choices, so we chose $0.01$ because the contrast maps more closely matched the contrast maps from the original data. We also tried several different resolutions for manually adding a grid of points at the two lowest frequencies. There were small changes in RMSPE for the different choices, so we chose a middle setting with 6,840 locations in the grid.

We also varied the number of principal components of the log time series spectra to include in the spatially-varying time series models. We show in Table \ref{pcsensitivity} that increasing the number of PCs beyond 1 increases the area-weighted RMSPE. This is likely because the models with fewer PCs require fewer time series parameters, which allows us to store more Fourier coefficients.

\begin{table} [H]
\centering
\begin{tabular}{ccccc}
 & & \multicolumn{3}{c}{Area-weighted RMSPE}\\
Ratio & number of PCs & Ocean & Land & Overall \\
\hline \hline
& 1 & 0.6937  &  0.3239  &  0.4618 \\
 & 2 & 0.7131 &  0.3319  &  0.4742  \\
20:1& 3 & 0.7355  &  0.3382  &  0.4871 \\
& 4 & 0.7647  &  0.3503  &  0.5058 \\
& 5 & 0.7940  &  0.3621  &  0.5243 \\
\hline
& 1 & 0.4202  &  0.2153  &  0.2896 \\
& 2 & 0.4276  &  0.2191  &  0.2947  \\
10:1& 3 & 0.4368 &   0.2207  &  0.2994 \\
& 4 & 0.4473  &   0.2258  &  0.3065 \\
& 5 & 0.4577  &  0.2298  &   0.3130 \\
\hline
& 1 & 0.1971  &  0.1225  &  0.1479 \\
 & 2 & 0.2006  &  0.1239  &  0.1501 \\
5:1& 3 & 0.2042 &   0.1252  &  0.1522 \\
& 4 & 0.2083  &  0.1272  &  0.1550 \\
& 5 & 0.2123 &   0.1288  &  0.1575 \\
\hline \hline
\end{tabular}
\caption{Area-weighted RMSPE for different numbers of time series principal components.}
\label{pcsensitivity}
\end{table}

In the greedy selection of stored coefficients, we added multiple coefficients simultaneously subject to the constraint that the locations of simultaneously  added coefficients could not be within distance $d_{\min}$ of one another. The results in Table \ref{dminsensitivity} indicate that the area-weighted RMSPE values have some dependence on $d_{\min}$. We chose $d_{\min} = 0.05$ in our analysis.

\begin{table} [H]
\centering
\begin{tabular}{ccccc}
 & & \multicolumn{3}{c}{Area-weighted RMSPE}\\
Ratio & $d_{\min}$ & Ocean & Land & Overall \\
\hline \hline
& 0.10 & 0.6962  &  0.3224 &   0.4622 \\
20:1 & 0.05 & 0.6937   & 0.3239  &  0.4618  \\
& 0.01 & 0.7490   & 0.3475  &  0.4975 \\
\hline
& 0.10 & 0.4228  &  0.2161  &  0.2911\\
10:1 & 0.05 & 0.4202  &  0.2153  &  0.2896  \\
& 0.01 & 0.4426  &  0.2273  &  0.3053 \\
\hline
& 0.10 & 0.2001 &   0.1233  &  0.1495\\
5:1 & 0.05 & 0.1971  &  0.1225  &  0.1479 \\
& 0.01 & 0.2043  &  0.1273  &  0.1535 \\
\hline \hline

\end{tabular}
\caption{Area-weighted RMSPE for different choices of minimum distance parameter $d_{\min}$.}
\label{dminsensitivity}
\end{table}

\bibliographystyle{apalike}
\bibliography{refs}

\end{document}